\documentclass[aap,preprint]{imsart}

\usepackage[OT1]{fontenc}
\usepackage{amsthm,amsmath}
\usepackage[numbers]{natbib}
\usepackage{url}
\usepackage[colorlinks,citecolor=blue,urlcolor=blue]{hyperref}
\usepackage[english]{babel}
\usepackage[utf8x]{inputenc}
\usepackage{amsmath,amssymb,hyperref}
\usepackage{graphicx}
\usepackage[colorinlistoftodos]{todonotes}
\usepackage[numbers]{natbib}

\usepackage{tikz-cd} 
\usepackage{commath}

\arxiv{arXiv:1707.08036}

\startlocaldefs
\newcommand{\N}{\mathcal{N}}
\newtheorem{theorem}{Theorem}

\newtheorem{lemma}[theorem]{Lemma}
\newtheorem{assumption}{Assumption}
\numberwithin{equation}{section}
\theoremstyle{plain}
\newcommand{\R}{\mathbb{R}}
\newcommand{\Rd}{\mathbb{R}^d}
\renewcommand{\P}{\mathbb{P}}
\newcommand{\E}{\mathbb{E}}

\endlocaldefs

\begin{document}

\begin{frontmatter}
\title{Theoretical properties of quasi-stationary Monte~Carlo methods}
\runtitle{Quasi-stationary Monte Carlo}

\begin{aug}
\author{\fnms{Andi Q.} \snm{Wang}\thanksref{t1,m1}\ead[label=e1]{a.wang@stats.ox.ac.uk}},

\author{\fnms{Martin} \snm{Kolb}\thanksref{m2}\ead[label=e2]{kolb@math.uni-paderborn.de}},
\author{\fnms{Gareth O.} \snm{Roberts}\thanksref{t2,m3}\ead[label=e3]{gareth.o.roberts@warwick.ac.uk}},
\and
\author{\fnms{David} \snm{Steinsaltz}\thanksref{m1}
\ead[label=e4]{steinsal@stats.ox.ac.uk}}

\thankstext{t1}{Supported by EPSRC OxWaSP CDT through grant EP/L016710/1.}
\thankstext{t2}{Supported by EPSRC grants EP/K034154/1, EP/K014463/1, EP/D002060/1.}
\runauthor{A. Wang et al.}

\affiliation{University of Oxford\thanksmark{m1}, Paderborn University\thanksmark{m2} and University of Warwick\thanksmark{m3}}

\address{A. Q. Wang\\
D. Steinsaltz\\
Department of Statistics\\
University of Oxford\\
24-29 St Giles\\
Oxford OX1 3LB\\
United Kingdom\\
\printead{e1}\\
\phantom{E-mail:\ }\printead*{e4}}

\address{M. Kolb\\
Department of Mathematics\\
Paderborn University\\
Warburger Str. 100\\
33098 Paderborn\\
Germany\\
\printead{e2}}

\address{G. O. Roberts\\
Department of Statistics\\
University of Warwick\\
Coventry CV4 7AL\\
United Kingdom\\
\printead{e3}}

\end{aug}

\begin{abstract}
\hspace{0.1mm} This paper gives foundational results for the application of quasi-stationarity to Monte Carlo inference problems. We prove natural sufficient conditions for the quasi-limiting distribution of a killed diffusion to coincide with a target density of interest. We also quantify the rate of convergence to quasi-stationarity by relating the killed diffusion to an appropriate Langevin diffusion. As an example, we consider in detail a killed Ornstein--Uhlenbeck process with Gaussian quasi-stationary distribution.
\end{abstract}

\begin{keyword}[class=MSC]
\kwd[Primary ]{60J60}
\kwd{60J70}
\kwd[; secondary ]{47D07}
\kwd{65C05}
\end{keyword}

\begin{keyword}
\kwd{Quasi-stationary Monte Carlo methods}
\kwd{quasi-stationary distributions}
\kwd{killed diffusions}
\kwd{$R$-theory}
\end{keyword}

\end{frontmatter}

\section{Introduction}
\label{sec:intro}

\subsection{Background} \label{sec:background}
\sloppy Markov chain Monte Carlo (MCMC) is a staple tool for statisticians wishing to perform Bayesian inference. Suppose we wish to sample approximately from the distribution $\pi$. The celebrated Metropolis--Hastings algorithm constructs an irreducible, aperiodic Markov chain $(Y_n)_{n=1}^\infty$ that is reversible with respect to $\pi$, hence has $\pi$ as its stationary distribution. General theory of Markov chains tells us that the distribution of $Y_n$ converges to $\pi$ as $n\rightarrow \infty$. 
The computations may, however, be intractable for large datasets and
high-dimensional models, such as modern `Big Data' applications often
demand: for a dataset of size $N$, merely evaluating the posterior distribution, of the form 
\begin{equation} \pi(x)\propto \prod_{i=1}^N f_i(x)\ , 
\label{eq:product}
\end{equation}
is an expensive $O(N)$ computation at each Markov chain iteration.

In \cite{Pollock2016}, the authors proposed the Scalable Langevin Exact (ScaLE) algorithm as part of a new Monte Carlo framework that is
provably efficient for Big-Data Bayesian inference. Starting with a diffusion $(X_t)_{t\ge 0}$ (in their case, a Brownian motion), a stopping time $\tau_\partial$, the ``killing time'', is defined in such a way that the \textit{quasi-limiting} distribution (sometimes termed the {\em Yaglom} limit) is $\pi$. That is, we have convergence of the conditional laws
\begin{equation}
\mathbb P_x(X_t \in \cdot \,|\tau_\partial >t) \rightarrow \pi(\cdot) \quad \text{ as }t\rightarrow \infty
\label{eq:intro_qsl}
\end{equation}
in an appropriate sense from any starting point $X_0 =x\in \mathbb R^d$. Such a $\pi$ is also \textit{quasi-stationary}, in the sense that 
\begin{equation}\mathbb P_\pi (X_t \in \cdot \,|\tau_\partial >t) = \pi(\cdot)
\label{eq:qs}
\end{equation}
for all $t\ge 0$, where $\P_\pi$ denotes the law of the process conditional on $X_0 \sim \pi$.
Any Monte Carlo procedure which aims to sample from a quasi-stationary distribution, for instance, using \eqref{eq:intro_qsl}, will be termed a {\em quasi-stationary Monte Carlo method}.

Quasi-stationarity has long been a subject of intensive study in the probability literature, summarised recently in \cite{Collet2012} and the bibliography of \cite{Pollett}. However, the ScaLE algorithm is the first application of quasi-limiting convergence to Monte Carlo sampling. Its attractiveness to the aforementioned `Big Data' problems stems from the fact that the ScaLE algorithm can be implemented in substantially less than $O(N)$ computing time (per unit stochastic process time). In fact, the algorithm is sometimes $O(1)$ and typically no worse than $O(\log (N))$. This is because the simulation of killed diffusions can be performed perfectly through \textit{subsampling} [usually using subsets of size $O(1)$] and, therefore, without any bias. Direct approaches based around subsampling a random subset of the $N$ terms in \eqref{eq:product} to obtain an estimate of the product have been proposed, although this results in unacceptably large errors in the target distribution unless the subset itself is $O(N)$; see, for instance, the discussions in \cite{Bardenet17}.

\citet{Pollock2016} gives some theory for the convergence properties of ScaLE, although this requires various regularity conditions which are difficult to check in many realistic statistical contexts. Our paper will give a much more complete picture under substantially weaker regularity conditions, and help to link quasi-stationary Monte Carlo methods with the established literature on quasi-stationarity.

Quasi-stationary convergence differs in important respects from the
more familiar theory of stationary convergence.
For a start, the theory of MCMC algorithms is most commonly formulated for discrete-time chains, whereas the ScaLE algorithm is fundamentally a continuous-time algorithm. There may be many probabilities $\pi $ which satisfy \eqref{eq:qs}, despite irreducibility, so we need to identify the
appropriate candidate for the limit \eqref{eq:intro_qsl}.
Perhaps most significant, the conditioned laws in \eqref{eq:intro_qsl} are not {\em consistent}: they are not the marginal laws of a single Markov process at time $t$. 
This prevents us from using much of the standard probabilistic armamentarium based on conditioning and the Markov property. Instead, to prove convergence we use R. Tweedie's $R$-theory, \cite{Tweedie1974}, and to study rates of convergence we follow the approach pioneered by \cite{Mandl1961}, drawing on the theory of semi-groups generated by linear differential operators.

\subsection{Summary of main results}
\label{sec:mainres}

We now summarise our main results, leaving the exact mathematical setting to be explicated in Section \ref{sec:back}. 
We will be assuming throughout the following.
\begin{assumption} \label{assump:L0}
$\pi$ is a positive, smooth and integrable function on $\mathbb R^d$.
\end{assumption}
Consider the $d$-dimensional diffusion $X=(X_t)_{t\ge 0}$, defined as the (weak) solution of the stochastic differential equation (SDE)
\begin{equation}
\dif X_t = \nabla A(X_t) \dif t + \dif W_t, \quad X_0 = x \in \mathbb R^d,
\label{eq:SDE_X}
\end{equation}
where $W$ is a standard $d$-dimensional Brownian motion and $\nabla$ denotes the gradient operator. We require the following.

\begin{assumption}
$A:\Rd \rightarrow \R$ is a smooth function such that the SDE (\ref{eq:SDE_X}) has a unique nonexplosive weak solution.
\label{assump:reg_conds}
\end{assumption}
Suppose we wish to sample from a distribution $\pi$ on $\Rd$ with a Lebesgue density, which we will also denote by $\pi$ --- the {\em target density} --- satisfying Assumption \ref{assump:L0}. We are typically thinking of applications in which
we have a statistical model and observed data for which $\pi$ is the Bayesian posterior distribution. We would like to construct a \textit{killing rate} $\kappa:\mathbb R^d \rightarrow [0,\infty)$
that makes $\pi$ into the quasi-limiting distribution of the diffusion $X$.
That is, we define the killing time 
\begin{equation}
\tau_\partial := \inf\bigg\{t\ge0: \int_0^t \kappa(X_s)\dif s>\xi\bigg \},
\label{eq:tau_d}
\end{equation}
where $\xi$ is an exponential random variable with parameter 1 independent of $X$. 
This killing time $\tau_\partial$, when the cumulative hazard function $t\mapsto \int_0^t \kappa(X_s)\dif s$ exceeds the (independent) threshold $\xi$, is equivalent to the first arrival time of a (doubly stochastic) Poisson process with rate function $t\mapsto \kappa(X_t)$.

We show that 
\begin{equation}
\begin{split}
\mathbb P_x(X_t \in E \,|\tau_\partial > t)&\rightarrow \pi(E) \text{ as }t\to\infty\\
\text{for all }x\in\Rd &\text{ and Borel-measurable }E\subset \Rd.
\end{split}
\label{eq:quasilimit}
\end{equation}

To have confidence that this convergence is practically meaningful for a sampling
algorithm, we need in addition to have some control over the rate of the convergence.

Our first result gives natural conditions under which the convergence (\ref{eq:quasilimit}) holds.

To begin with, we require the following compatibility condition between the tails of $\pi$ and the underlying diffusion.
\begin{assumption}
\begin{equation*}
\int_{\mathbb R^d}\frac{\pi^2(y)}{\exp(2A(y))}  \dif y<\infty.
\end{equation*}
\label{assump:L2}
\end{assumption}
Assumption \ref{assump:L2} is natural from a statistical point of view. Recall that without killing, the diffusion $X$ has invariant density proportional to $\exp(2A)$ if this quantity is integrable (and certain regularity conditions hold; see \cite{Roberts1996}, Theorem 2.1). Assumption \ref{assump:L2} can then be interpreted as requiring that the likelihood ratio $\pi(Y)/e^{2A(Y)}$ has finite variance when $Y\sim \exp(2A)$. This is what we would need to assume were we to target $\pi$ by importance sampling from $\exp(2A)$.

In particular, Assumption \ref{assump:L2} holds when the stronger `rejection sampling' condition holds: that there exists some $M<\infty$ such that
\begin{equation}
\frac{\pi(y)}{\exp(2A(y))}<M \quad \forall y \in \mathbb R^d.
\label{eq:reject_sampl}
\end{equation}
If $\exp(2A)$ is integrable, then this is precisely the condition that would allow us to sample from $\pi$ using a rejection sampler with proposal density proportional to $\exp(2A)$. Informally, this demands that the asymptotic tail behavior of the diffusion be heavier than the tails of the target distribution. In particular, if the diffusion $X$ is a Brownian motion on $\mathbb R^d$ ($A\equiv 0$ in \eqref{eq:SDE_X}), Assumption \ref{assump:L2} holds whenever the target density $\pi$ is bounded.

We now define the appropriate killing rate $\kappa$, to be used to construct the killing time $\tau_\partial$ in \eqref{eq:tau_d}. Define $\tilde \kappa:\mathbb R^d \rightarrow \mathbb R$ by
\begin{equation}
\tilde \kappa (y) := \frac{1}{2}\bigg (\frac{\Delta \pi}{\pi} - \frac{2\nabla A \cdot \nabla \pi}{\pi} - 2\Delta A \bigg)(y), \quad y \in \mathbb R^d
\label{eq:kappa_tilde}
\end{equation}
where $\Delta$ denotes the Laplacian operator. We require the following.
\begin{assumption}
$\tilde \kappa$ is bounded below, and not identically zero.
\label{assump:kappa}
\end{assumption}
We will see that the correct killing rate is 
\begin{eqnarray}
\kappa =\tilde\kappa+K,
\label{eq:kappa}
\end{eqnarray}
where $K:=-\inf_{y\in \mathbb R^d} \tilde \kappa(y)$, chosen so that $\kappa$ is nonnegative everywhere. If $\tilde \kappa$ is identically zero, then there is no killing and we are in the familiar realm of stationary convergence of (unkilled) Markov processes; in fact, $X$ will be a Langevin diffusion targeting $\pi$; see \cite{Roberts1996}. To facilitate the development of intuition, some examples of $\kappa$ in the case of $A\equiv 0$ are given in Section \ref{subsec:eg_kappa}.
Heuristically, this form for the killing rate makes $\pi$ an eigenfunction for the generator of the killed diffusion, which corresponds to quasi-stationarity; see Section \ref{sec:back} for the mathematical details and further explanation.

The form of the untranslated killing rate in \eqref{eq:kappa_tilde} also has the natural following interpretation. Writing $U:=\log \pi$, which we can do since we are assuming $\pi$ is positive, and as above thinking of $\exp(2A)$ as describing the asymptotic unkilled dynamics, we can rewrite \eqref{eq:kappa_tilde} as
\begin{equation}
\tilde \kappa(y)= \frac{1}{2} \big ( \Delta (U-2A) + \nabla U\cdot \nabla(U - 2 A) \big)(y), \quad y\in \Rd.
\label{eq:tkappa_logs}
\end{equation}
Written this way, we see $\tilde \kappa$ is a measure of the discrepancy between the derivatives of $\log \pi$ and $2A$, and Assumption \ref{assump:kappa} states that this discrepancy cannot be arbitrarily negative.

\subsection{Convergence to quasi-stationarity}
\begin{theorem}
Suppose Assumptions \ref{assump:L0}, \ref{assump:reg_conds}, \ref{assump:kappa} and \ref{assump:L2} and  hold. Then $X$ has quasi-limiting distribution $\pi$. That is, the convergence in \eqref{eq:quasilimit} holds.
\label{thm:qsd}
\end{theorem}

\noindent\textbf{Remarks}
\begin{enumerate}
\item This significantly improves on Theorem 1 of \cite{Pollock2016}: their result only applied to killed Brownian motions, and their complicated condition on the tails of the target density has been removed. While Brownian motion---$A\equiv 0$ in \eqref{eq:SDE_X}---is a natural choice of a `proposal' diffusion, with developments in the exact simulation of diffusions, such as \cite{Beskos2006}, there is potential to consider other diffusions as candidates. In Section \ref{sec:OU_example}, we consider an Ornstein--Uhlenbeck process targeting a Gaussian distribution.
\item We are not able to use the recent convergence results of \cite{Champagnat2016a}. Their approach is via minorisation-type conditions, which do not hold in our particular noncompact state space setting, and so we cannot apply their theorem on uniform exponential convergence.
\item Assumption \ref{assump:L2} is in fact not a necessary condition. For example in Section 4.6 of \cite{Kolb2012} the authors consider cases of low killing on $[0,\infty)$, where $\lambda_0^\kappa$, the bottom of the spectrum (in our case $K$; see Section \ref{sec:background_operator}), is not an eigenvalue in the $\mathcal L^2$ sense, but convergence to quasi-stationarity still occurs. Instead, the requirement is that the unkilled process be recurrent. In the context of quasi-stationary Monte Carlo methods, where we are free to choose the diffusion, Assumption \ref{assump:L2} is a natural condition, since the excluded cases have zero spectral gap, hence inevitably poor convergence properties.
\item Theorem \ref{thm:qsd} also extends the results of \cite{Kolb2012}: there the authors considered only (one-dimensional) cases where $\lim_{y\rightarrow \infty}\kappa(y) \neq \lambda_0^\kappa$. For example, our result gives convergence of killed Brownian motions with polynomially-tailed quasi-stationary distributions: in such cases 
$$\lim_{\|y\|\rightarrow \infty} \kappa(y)= \lambda_0^\kappa,$$
but the conditions of Theorem \ref{thm:qsd} still hold, so we obtain convergence to quasi-stationarity.
\item We also obtain convergence of the conditional measures $\P_x (X_t \in \cdot \,|\tau_\partial >t)$ to $\pi$ in total variation distance as $t\rightarrow \infty$, as shown in the proof of Theorem 7 of \cite{Tuominen1979}.
\end{enumerate}

\subsection{Rate of convergence}
Our second result helps us to understand the rate of convergence to quasi-stationarity. Let $Z=(Z_t)_{t\ge 0}$ be the weak solution of the related SDE
\begin{equation}
\dif Z_t = \frac{1}{2}\nabla \log\bigg (\frac{\pi^2}{\exp(2A)}\bigg )(Z_t)\dif t + \dif W_t,
\label{eq:Lang_pi2gamma}
\end{equation}
with $Z_0=x$. This is an example of a Langevin diffusion. Under suitable regularity conditions (see Theorem 2.1 of \cite{Roberts1996}), the law of the diffusion $Z_t$ converges to the distribution on $\Rd$ with Lebesgue density proportional to $\pi^2/\exp(2A)$ as $t\rightarrow \infty$. (Assumption \ref{assump:L2} guarantees that this is integrable.) Let $-L^{Z}$ denote the infinitesimal generator of this process and let $-L^{\kappa}$ denote the infinitesimal generator of the process (\ref{eq:SDE_X}) killed at rate $\kappa$. These operators will be constructed explicitly in Section \ref{sec:background_operator} as self-adjoint operators on the appropriate $\mathcal L^2$ Hilbert spaces.

Writing $\gamma:= \exp(2A)$, $\Gamma(\dif x):= \gamma(x)\dif x$ for the corresponding Borel measure on $\Rd$, which is the reversing measure of the diffusion $X$, and $\varphi := \pi/\gamma$, we have the following result.

\begin{theorem}
Under the same conditions as Theorem \ref{thm:qsd}, the $\mathcal L^2$ spectra of $L^{Z}$ and $L^{\kappa}$ agree, up to an additive constant. In particular, when $L^Z$ has a spectral gap, the transition kernel of the killed process $p^\kappa(t,x,y)$ satisfies
$$
\bigl| e^{tK}p^\kappa(t,x,y)- \varphi(x)\varphi(y) \bigr| \le C e^{-t(\lambda_1^Z-\lambda_0^Z)},
$$
where $\lambda_1^Z>\lambda_0^Z=0$ are the bottom two eigenvalues of the Langevin diffusion, and the constant $C$ may depend on $x$ and $y$. If the drift
in \eqref{eq:SDE_X} is bounded then $C$ may be chosen independent of $x$ and $y$.

If the measure $\Gamma$ is such that $\Gamma(\R^d)<\infty,$ then for an initial $\Gamma$-density $\psi \in \mathcal L^1(\Gamma)\cap \mathcal L^2(\Gamma)$,
$$
  \bigl|\mathbb P_\psi (X_t \in E \, |\, \tau_\partial >t) -\pi(E)\bigr|
\le C' e^{-t(\lambda_1^Z-\lambda_0^Z)},
$$
for any measurable $E\subset \Rd$, where
$$
C'= 
\frac{2\left(\int \psi(x)^2\dif\Gamma(x)\right)^{1/2}\Gamma\left( \R^d \right)^{1/2}}{\int \psi(x)\pi(x)\dif x\cdot\int \pi(x)\dif x}.
$$
\label{thm:rate_conv}
\end{theorem}
The additive constant in Theorem \ref{thm:rate_conv} is $K$; that is, the spectrum of $L^{Z}$ is the
translation of the spectrum of $L^{\kappa}$ by $+K$.

Theorem \ref{thm:rate_conv} tells us that the stationary convergence of the Langevin diffusion \eqref{eq:Lang_pi2gamma} and the quasi-stationary convergence of our killed diffusion occur at the same exponential rate, given by the equal spectral gaps. Since Langevin dynamics have been applied widely in computational statistics and the applied sciences, their rates of convergence have been studied extensively; see, for instance, the recent results of \cite{Dalalyan2016} and \cite{Durmus2015}. Thus for many cases of $\pi$ we will be able to accurately describe the rate of convergence in (\ref{eq:quasilimit}).

Theorem \ref{thm:rate_conv} also suggests that quasi-stationary Monte Carlo methods relying on \eqref{eq:quasilimit} may converge relatively slowly for densities which are \textit{multimodal}. In the case of $A\equiv 0$ (killed Brownian motion), if $\pi$ is multimodal, then $\pi^2$ will typically be even more irregular, and the Langevin diffusion targeting $\pi^2$ will converge only gradually. On the other hand, quasi-stationary Monte Carlo methods should have good success targeting densities which are unimodal, such as logconcave densities. If $\pi$ is unimodal, $\pi^2$ will be even more regular and have faster tail decay, leading to faster convergence of the Langevin diffusion. Such densities appear naturally in the context of Big-Data Bayesian inference. The Bernstein--von Mises theorem (\cite[Section 10.2]{Vaart2000}) tells us, for instance, that for large datasets
the posterior distributions are approximately Gaussian.\\[4mm]

\noindent\textbf{Remarks} 
\begin{enumerate}
\item A sufficient condition for the existence of a spectral gap ($\lambda_1^\kappa > \lambda_0^\kappa$) is that
\begin{equation}
\liminf_{\|x\|\rightarrow \infty} \tilde \kappa(x) > 0.
\label{eq:cond_spec_gap}
\end{equation}
See, for instance, the proof of Lemma 3.3(v) of \cite{Kolb2012}, which carries over into our setting. Furthermore, if $\liminf_{\|x\|\rightarrow \infty} \tilde \kappa(x)=+\infty$ then this implies that the spectrum is purely discrete (the \textit{essential spectrum} is empty). In the case of killed Brownian motion this holds for all exponentially-tailed densities of the form $\exp(-\beta\|x\|^\alpha )$ for some $\beta>0, \alpha\ge 1$.
\item The Langevin diffusion in (\ref{eq:Lang_pi2gamma}) is precisely the $Q$-process
(the diffusion conditioned never to be killed) defined by the diffusion $X$ and 
the killing time $\tau_\partial$. It is defined as the limit 
\begin{equation*}
\mathbb Q_x(A):=\lim_{T\rightarrow \infty} \mathbb P_x (A | T<\tau_\partial)
\end{equation*}
for $A\in \sigma(X_s:s\le t)$ for some $t\ge 0$.
\item Theorem \ref{thm:rate_conv} is a continuous state-space generalisation of Theorem 1 of \cite{Diaconis2014}: there the authors showed that in a finite state-space, rates of convergence to quasi-stationarity in total variation distance can be bounded above and below by constant multiples of the rates of convergence to stationarity in total variation of an appropriate unkilled process. 

\end{enumerate}

\subsection{Examples of $\kappa$}
\label{subsec:eg_kappa}
In the simple and computationally important case of a killed Brownian motion ($A\equiv 0$ in \eqref{eq:SDE_X}), $\tilde \kappa$ as defined in \eqref{eq:kappa_tilde} simplifies down to
\begin{equation*}
\tilde \kappa(y) = \frac{\Delta\pi}{2\pi},\quad y\in\Rd.
\end{equation*}
In the following examples it can be easily checked that the conditions of Theorem \ref{thm:qsd} are satisfied.
\begin{itemize}
\item \textit{Gaussian on $\Rd$.} Let $\sigma^2>0$ and $\pi(y)\propto \exp(-\|y\|^2/(2\sigma^2))$ for $y\in \Rd$, where throughout $\|\cdot\|$ denotes the Euclidean norm. Then straightforward calculation gives us that $\tilde \kappa(y)=\frac{1}{2}(\sigma^{-4}\|y\|^2-\sigma^{-2}d)$ for $y\in\Rd$ and hence
\begin{equation*}
\kappa(y)=\frac{1}{2\sigma^4}\|y\|^2, \quad y\in\Rd.
\end{equation*}
Since $\liminf_{\|y\|\to\infty}\tilde\kappa(y)>0$ (in fact it's infinite), we expect exponential rates of convergence to quasi-stationarity, from condition \eqref{eq:cond_spec_gap}. This example is considered in some detail in the case $d=1$ in Section \ref{sec:OU_example}. This example also gives the independently interesting result that a Brownian motion on $\Rd$ killed at a quadratic rate will have a Gaussian quasi-limiting distribution.

\item \textit{Univariate exponential decay.} Consider a one-dimensional, positive, smooth target density $\pi$ with tail decay $\pi\propto \exp(-\beta |y|)$ for all $y$ outside of a compact set $E\subset \R$, for some $\beta>0$. We find that for all $y \in \R \backslash E$, $\tilde \kappa(y) = \beta^2$, that is, a positive constant. The killing rate $\kappa$ will then also be constant asymptotically. By condition \eqref{eq:cond_spec_gap}, we expect exponential convergence to quasi-stationarity.

\item \textit{Heavy-tailed case.} Consider a univariate Cauchy target, $\pi(y)\propto 1/(1+y^2)$ for $y\in\R$. Then simple calculation gives $\tilde \kappa(y)=\frac{3y^2-1}{(1+y^2)^2}$, for $y\in\R$ and then $$\kappa(y)=\frac{3y^2-1}{(1+y^2)^2}+1, \quad y\in \Rd.$$ We see here an example where $\liminf_{|y|\to\infty}\tilde \kappa(y) = 0$; the sufficient condition for a spectral gap \eqref{eq:cond_spec_gap} fails and we expect slower convergence.
\end{itemize}

\section{Example: Ornstein--Uhlenbeck process targeting a Gaussian density}
\label{sec:OU_example}

Before turning to the mathematical technicalities, we offer a mathematically tractable example that can be readily simulated: a killed Ornstein--Uhlenbeck process targeting a Gaussian distribution. For simplicity of presentation we discuss the univariate case $d=1$. Analogous results hold in the multivariate case, but the notation is more cumbersome, and the calculations more involved.

Throughout this section, we write $\N(\mu, \sigma^2)$ with $\mu \in \mathbb R, \sigma^2>0$ to denote the univariate Gaussian distribution with mean $\mu$ and variance $\sigma^2$.

In \eqref{eq:SDE_X}, we let $A(y)=-(\nu-y)^2 /(4\tau^2)$ for each $y\in \mathbb R$, where $\nu\in \mathbb R, \tau^2>0$ are fixed. This defines a diffusion $X$ as the weak solution of
\begin{equation}
\dif X_t = \frac{1}{2\tau^2}(\nu-X_t) \dif t + \dif W_t, \quad X_0=x.
\label{eq:OU_process}
\end{equation}
The Ornstein--Uhlenbeck process $X$ has a $\N(\nu, \tau^2)$ stationary distribution; the corresponding density function is proportional to $\exp(2A)$.

Fix $\mu\in \mathbb{ R}$ and $\sigma^2>0$, and let the target density be $$\pi(y) = \exp\bigg \{-\frac{1}{2\sigma^2}(y-\mu)^2\bigg \}$$ for each $y\in \mathbb R$, the (unnormalised) density of a $\N(\mu, \sigma^2)$ random variable. We note that the regularity conditions---Assumptions \ref{assump:L0} and \ref{assump:reg_conds}---hold.

The untranslated killing rate computed from \eqref{eq:kappa_tilde} is for each $y\in \R$ given by
\begin{equation}
\tilde \kappa(y) = \frac{1}{2}\bigg(\frac{(y-\mu)^2}{\sigma^4}-\frac{1}{\sigma^2}+\frac{(\nu-y)(y-\mu)}{\tau^2 \sigma^2}+ \frac{1}{\tau^2}\bigg).
\label{eq:OU_kappatilde}
\end{equation}
We now assume
\begin{equation}
\tau^2>\sigma^2;
\label{eq:OU_tail_var}
\end{equation}
that is, the invariant distribution of the underlying diffusion has tails that are heavier than those of the target distribution. 
This makes the leading coefficient in the quadratic \eqref{eq:OU_kappatilde} positive, so that $\tilde\kappa$
is bounded below, meaning that Assumption \ref{assump:kappa} holds.
In this case, we will have a spectral gap (since the limit of the killing at infinity is $+\infty$; see \eqref{eq:cond_spec_gap}, so we expect quasi-stationary
convergence to occur at an exponential rate. Completing the square in \eqref{eq:OU_kappatilde} gives the minimum value
\begin{equation}
K:= -\inf_{y\in \mathbb R} \tilde \kappa(y)=\frac{(\mu-\nu)^2}{8\tau^2(\tau^2-\sigma^2)} + \frac{\tau^2-\sigma^2}{2\tau^2\sigma^2}.
\label{eq:OU_lambda_0}
\end{equation}
In Section \ref{sec:background_operator} we will identify $K$ with $\lambda_0^\kappa$, the bottom of the $\mathcal L^2$-spectrum of the generator of the killed diffusion, and so $K$ is also the asymptotic rate of killing (see Lemma 4.2 of \cite{Kolb2012}). We see from \eqref{eq:OU_lambda_0} that $K$ is strictly positive, as our calculation in Section \ref{sec:background_semi-group} predicts. Adding $K$ to $\tilde\kappa$ and rearranging, we obtain the killing rate
\begin{equation}
\kappa(y)=\frac{\tau^2 - \sigma^2}{2\tau^2 \sigma^4} \bigg( y-\bigg\{ \frac{\mu+\nu}{2}+\frac{\tau^2}{\tau^2-\sigma^2}\bigg(\mu-\frac{\mu+\nu}{2}\bigg)\bigg\}\bigg)^2
\label{eq:OU_kappa}
\end{equation}
for $y\in \mathbb R$.

It remains to check Assumption \ref{assump:L2}. By direct calculation
\begin{equation*}
\frac{\pi^2}{\exp(2A)} (y) \propto \exp\bigg\{-\frac{1}{2} \frac{2\tau^2-\sigma^2}{\sigma^2\tau^2} \bigg(y-\frac{2\mu\tau^2 -\nu\sigma^2}{2\tau^2 -\sigma^2}\bigg)^2\bigg\}.
\end{equation*}
Our assumption \eqref{eq:OU_tail_var} guarantees this will be integrable, and in fact proportional to the density of the Gaussian distribution
\begin{equation}
\N\bigg(\frac{2\mu\tau^2-\nu\sigma^2}{2\tau^2 -\sigma^2},\frac{\sigma^2 \tau^2}{2\tau^2-\sigma^2}\bigg).
\label{eq:OU_pi2gamma_dens}
\end{equation}
So Theorem \ref{thm:qsd} allows us to conclude that $\pi$ is 
the quasi-limiting distribution of our Ornstein--Uhlenbeck process (\ref{eq:OU_process}) killed at rate (\ref{eq:OU_kappa}), as long as (\ref{eq:OU_tail_var}) holds.

Since $\pi^2 /\exp(2A)$ is the density of a Gaussian distribution, it follows that the corresponding Langevin diffusion (\ref{eq:Lang_pi2gamma}) is another Ornstein--Uhlenbeck process, albeit with stationary distribution given by (\ref{eq:OU_pi2gamma_dens}). In \cite{Metafune2002}, the authors explicitly computed the $\mathcal L^p$ spectra of Ornstein--Uhlenbeck operators, and by applying their Theorem 3.1 we find that the $\mathcal L^2$ spectrum of $L^Z$ is given by
\begin{equation*}
\Sigma\big (L^Z\big ) = \bigg\{ \lambda_n^Z=\frac{n(2\tau^2 - \sigma^2)}{2\sigma^2 \tau^2}: n=0,1,2,\dots\bigg \}.
\end{equation*}
By Theorem \ref{thm:rate_conv}, this coincides (up to an additive constant) with the spectrum of our killed process (\ref{eq:OU_process}). In particular, the spectral gap of our killed process is 
\begin{equation*}
\lambda_1^Z - \lambda_0^Z=\frac{2\tau^2-\sigma^2}{2\sigma^2 \tau^2}=\frac{1}{\sigma^2}-\frac{1}{2\tau^2}\; .
\end{equation*}
For this example, there are two mechanisms influencing the convergence to quasi-stationarity: the drift of the underlying diffusion \eqref{eq:OU_process}, along with the killing \eqref{eq:OU_kappa} and subsequent conditioning on survival. It is interesting to note that the spectral gap is maximised when $\tau^2\rightarrow \infty$, in which case the drift is 0. When in addition $\mu=\nu$, we see that the killing is also maximal, as measured by, say, the asymptotic killing rate \eqref{eq:OU_lambda_0}. This limit case $\tau^2 \rightarrow \infty$ corresponds to the case of killed Brownian motion [$A\equiv 0$ in \eqref{eq:SDE_X}]. This suggests that the rate of convergence to quasi-stationarity is determined more by the killing mechanism than by the underlying drift. However, depending on the method of implementation, a greater rate of killing could lead to reduced computational efficiency.

\begin{figure}
\centering
\includegraphics[width=\textwidth]{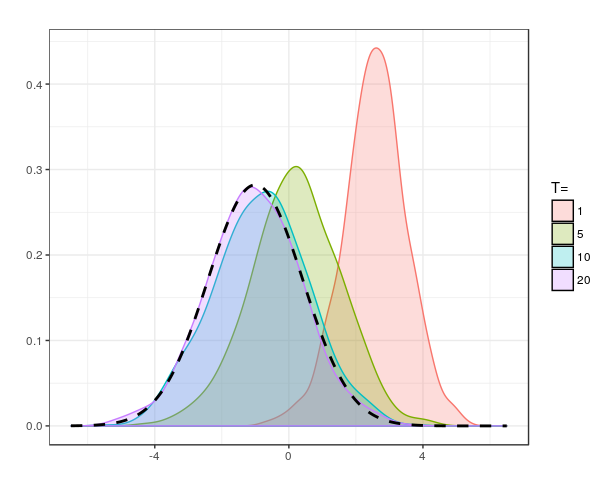}
\caption{Estimates of the conditioned laws $\mathbb P_x(X_T \in \cdot\,|\tau_\partial>T)$ for the process (\ref{eq:OU_process}) with parameters $\nu=2,\tau^2=4$ started at $x=3$ for various $T$. The dashed black line shows the quasi-stationary density, a Gaussian density with mean and variance $\mu=-1,\sigma^2=2,$ respectively. }
\label{fig:QSD_OU}
\end{figure}

This simple example is amenable to simulation, as shown in Figure \ref{fig:QSD_OU}. The figure shows the conditional distributions $\mathbb P_x (X_T \in \cdot \, |\tau_\partial > T)$ for $T=1,5,10,20$ for the choices $\nu=2, \tau^2=4, \mu=-1,\sigma^2=2$, and initial value $X_0=x=3$.

\section{Mathematical preliminaries}
\label{sec:back}

\subsection{Definitions}  \label{sec:background_def}
Fix $d\in \mathbb N$.  Let $\nabla$ denote the gradient operator; the $d$-dimensional vector with components $\nabla_i = \partial/\partial x_i$, $i=1,\dots,d$. We will denote the Laplacian operator by $\Delta:=\sum_{i=1}^d \pd[2]{\phantom{x}}{x_i}$.
We are given functions $A:\mathbb R^d \rightarrow\mathbb R$ and $\pi:\mathbb R^d \rightarrow [0,\infty)$
that satisfy Assumptions \ref{assump:L0}, \ref{assump:reg_conds} and \ref{assump:L2}.
For brevity we write 
$$
\gamma:=\exp(2A).
$$ 
$\pi$ is our target density, which need not be normalised. In a slight abuse of notation we will also write $\pi$ for the Borel probability measure on $\mathbb R^d$ with Lebesgue density proportional to $\pi$. 

Let $C\equiv C([0,\infty),\mathbb R^d)$ denote the space of continuous functions mapping $[0,\infty)\rightarrow\mathbb R^d$, and let $\omega$ be a typical element. For each $t\ge 0$, let $X_t: C\rightarrow \mathbb R^d$ be the coordinate mapping $X_t(\omega)=\omega(t)$, and let $\mathcal C:= \sigma(\{X_t: t \ge 0\})$ be the cylinder $\sigma$-algebra. For any $x\in\mathbb R^d$, let $\tilde{ \mathbb P}_x$ be the measure on $(C, \mathcal C)$ such that under $ \tilde {\mathbb P}_x$, $X=(X_t)_{t\ge 0}$ is the weak solution to (\ref{eq:SDE_X}).

Define $\kappa:\mathbb R^d \rightarrow [0,\infty)$ by
\begin{equation*}
\kappa(y):= \tilde \kappa(y) +K, \quad \forall y \in \mathbb R^d,
\end{equation*}
where $\tilde \kappa$, defined in (\ref{eq:kappa_tilde}), is required to
satisfy Assumption \ref{assump:kappa}, so that $K:=-\inf_{y\in \mathbb R^d} \tilde \kappa(y)$
is finite.
We augment our probability space to include an independent unit exponential random variable $\xi$,
and define killing at rate $\kappa$ as in \eqref{eq:tau_d}, denoting this augmented space by $(\Omega, \mathcal F, \mathbb P_x)$.

We define $\mathcal L^2(\Gamma)\equiv\mathcal L^2(\mathbb R^d, \Gamma)$ to be the Hilbert space of (equivalence classes of) Borel-measurable square-integrable functions $f,g: \mathbb R^d \rightarrow \mathbb R$ with respect to the inner product
\begin{equation*}
\langle f,g\rangle_{\mathcal L^2(\Gamma)} = \int_{\mathbb R^d} f(y) g(y) \,\dif\Gamma(y)
\end{equation*}
where the measure $\Gamma$ is given by $\dif \Gamma(y)=\gamma(y)\dif y$, with $\dif y$  denoting Lebesgue measure on $\Rd$. We denote the corresponding norm by $\|\cdot\|_{\mathcal L^2(\Gamma)}$.

Define $\varphi: \mathbb R^d \rightarrow\mathbb R$ by
\begin{equation*}
\varphi:=\frac{\pi}{\exp(2A)},
\end{equation*}
which is smooth and positive. By construction, we have that $\varphi$ is integrable with respect to $\Gamma$: 
\begin{equation*}
\int_{\Rd} \varphi(y) \dif\Gamma(y)=\int_{\Rd}\pi(y)\dif y <\infty.
\end{equation*}
We 
will generally be working in the function space $\mathcal L^2(\Gamma)$, as this is the space on which the generator of the killed diffusion can be realised as a self-adjoint operator, which we will do explicitly in Section \ref{sec:background_operator}. As such, we will want consider densities with respect to $\Gamma$---rather than Lebesgue measure---and hence we will work with $\varphi$, rather than directly with $\pi$. Of course in the case of killed Brownian motion, $A\equiv 0$, $\pi$ and $\varphi$ coincide.

Following this line of thought, Assumption \ref{assump:L2} states that indeed $\varphi \in \mathcal L^2(\Gamma)$:
$$
\|\varphi\|_{\mathcal L^2(\Gamma)}^2 = \int_{\mathbb R^d}\frac{\pi^2(y)}{\exp(2A(y))}  \dif y<\infty.
$$
Without loss of generality we can rescale $\pi$ so that this quantity is 1.

\subsection{The killed Markov semi-group} \label{sec:background_semi-group}
Our results depend on the spectral theory of self-adjoint linear operators on Hilbert spaces. The proof of Theorem \ref{thm:qsd} avoids the heavy machinery of this
theory by drawing on R. Tweedie's {\em R-theory}, which provides some of the
results of operator theory most relevant to asymptotics of stochastic processes
in a somewhat probabilistic package. We review the essentials of operator theory in Section
\ref{sec:background_operator}, but it will be required only for the proof of Theorem \ref{thm:rate_conv}.

The diffusion $X$ killed at rate $\kappa$ has a formal infinitesimal generator $-\tilde L^\kappa$ described by
\begin{equation}
\tilde L^\kappa = -\frac{1}{2 \exp(2A)} \nabla\cdot \exp(2A) \nabla +\kappa=-\frac{1}{2}\Delta -\nabla A \cdot \nabla + \kappa.
\label{eq:tilde_L_kappa}
\end{equation}
Under Assumption \ref{assump:kappa}, this formal differential operator can be realized as a positive self-adjoint operator $L^\kappa$ on an $\mathcal L^2$ Hilbert space. 
It is this theory that we defer to Section \ref{sec:background_operator}.

Straightforward calculation shows that
\begin{equation}
\tilde L^\kappa \varphi = K \varphi.
\label{eq:L_varphi}
\end{equation}
So $\varphi$ is an eigenfunction of the formal differential operator $\tilde L^\kappa$ with eigenvalue $K$. Since we have assumed that $\varphi$ is in $\mathcal{L}^2(\Gamma)$,
\begin{align*}
  K\int &\varphi(y)^2 \dif\Gamma(y)=
  \int \varphi(y) \tilde L^\kappa\varphi (y) \dif\Gamma(y)\\
   &= -\frac12 \int \varphi(y)\bigl(\Delta \varphi(y) + 2\nabla A \cdot \nabla \varphi(y) \bigr) \dif\Gamma(y)
     + \int \kappa(y)\varphi(y)^2 \dif\Gamma(y)\\
   &\ge \int \kappa(y)\varphi(y)^2 \dif\Gamma(y)\\
   &>0.
\end{align*}
The first inequality follows since $\varphi$ is smooth, so an application of Green's identity shows that the integral term is nonnegative. The final strict inequality follows since $\varphi$ and $\gamma$ are strictly positive and
$\kappa$ is not identically 0, by Assumption \ref{assump:kappa}. Thus we conclude $K>0$.

Recall from \cite{Tuominen1979} that a finite nonnegative measurable function $f$ with ${\int f (x)\dif x >0}$ is said to be 
\textit{$\lambda$-invariant} for a continuous-time semi-group $(P_t)_{t\ge 0}$ if for all $t>0$, 
$$ 
f(x)=e^{\lambda t}P_t f(x)\quad \text{ for almost every }x
$$ 
and a $\sigma$-finite nontrivial measure $\nu$ is \textit{$\lambda$-invariant} for continuous-time $(P_t)_{t\ge 0}$ if for all $t>0$, 
$$ \nu(A)=e^{\lambda t}\nu P_t(A) \text{ for every measurable }A.$$
Analogous notions of $R$-invariance of functions and measures are similarly defined for discrete-time processes as well; the requirement $t>0$ is replaced with $t\in \mathbb N$, and $e^{\lambda t}$ is replaced is replaced by $R^t$.

All we need
for present purposes is the following lemma.
\begin{lemma}
The sub-Markovian semi-group $(P^\kappa_t)_{t\ge 0}$of the killed process $X$ has a unique self-adjoint generator
that is an extension of $-\tilde L^\kappa$ on smooth compactly supported functions. $\pi$ is a $\lambda$-invariant measure for this semi-group, and $\varphi$ a $\lambda$-invariant function, for $\lambda=K$.
\end{lemma}

Except for some technical complications, which we will describe in the context of
presenting the operator-theory framework in Section \ref{sec:background_operator},
this should be reasonably intuitive. 
We have already pointed out in \eqref{eq:L_varphi}
that $\varphi$ is an eigenfunction of the generator with eigenvalue $-K$. Direct calculation
shows that $\tilde L^\kappa$
is symmetric with respect to the measure $\Gamma$; that is, for $f,g \in \mathcal{L}^2(\Gamma)$ in the domain of $\tilde L^\kappa$ we have that 
$$ \langle \tilde L^\kappa f,g\rangle_{\mathcal L^2(\Gamma)}=\langle f, \tilde L^\kappa g\rangle_{\mathcal L^2(\Gamma)}.$$ 
Heuristically, since our assumptions ensure that the generator of the killed diffusion $L^\kappa$ is symmetric, using \eqref{eq:L_varphi} we obtain the following manipulations, for any nonnegative test function $f\in \mathcal L^2(\Gamma)$:
\begin{align*}
\E_\pi[L^\kappa f(Y)]&=\int \pi(y) L^\kappa f(y)\dif y=\int \varphi(y)L^\kappa f(y)\dif\Gamma(y)
\\ &=\int L^\kappa \varphi(y)f(y)\dif \Gamma(y)=K\int \varphi(y)f(y)\dif \Gamma(y)
\\ &= K\E_\pi[f(Y)].
\end{align*}
Bearing in mind that $L^\kappa$ is \textit{minus} the generator of the killed diffusion, this shows that started in $\pi$ the process will remain in $\pi$, except with a \textit{mass loss at rate} $K$. That is to say, $\pi$ is quasi-stationary. For an unkilled diffusion, if $\pi$ were stationary, we would expect a similar expression to hold for any appropriate $f$, except with the right-hand side being exactly zero, reflecting the fact that the mass is preserved.

If we think of the
adjoint operator---acting on measures---as acting on densities with
respect to $\Gamma$, we have $(P^\kappa_t)^*g= (P^\kappa_t)g$. On the other hand, if $g$ is
a density with respect to Lebesgue measure the action is
\begin{equation} 
\label{eq:adjointaction}
P^\kappa_t g= \gamma P^\kappa_t(g/\gamma).
\end{equation}

\subsection{Operator theory} \label{sec:background_operator}
This section gives the mathematical background necessary for the proof of Theorem \ref{thm:rate_conv} in Section \ref{sec:rates_of_conv}. Readers interested in the proof of Theorem \ref{thm:qsd} can move straight to Section \ref{sec:proofqsd}.

Our operator $\tilde L^\kappa$ on $C^{\infty}_c(\mathbb{R}^d)$, smooth compactly supported functions, is a symmetric semi-bounded operator and, therefore, has a self-adjoint extension, for instance the Friedrichs extension; see \cite[Section 4.4]{davies1995spectral}.
As a matter of fact, our operator is essentially self-adjoint---proven in Section \ref{subsec:pf_thm2}---and thus has a unique self-adjoint extension $L^\kappa$, so its completions are self-adjoint.

Recall that \eqref{eq:tilde_L_kappa} describes the formal infinitesimal generator of our killed process. This gives rise to a closable densely-defined positive quadratic form $\tilde q^\kappa$ on $\mathcal L^2(\Gamma)$ given by
\begin{equation*}
\tilde q^\kappa(f) = \frac{1}{2} \int_{\mathbb R^d} \nabla f \cdot \nabla f (y) \gamma(y)\dif y + \int_{\mathbb R^d} \kappa(y) |f(y)|^2 \gamma(y)\dif y
\end{equation*}
for $f \in \mathcal D_\kappa$, where
\begin{equation*}
\mathcal D_\kappa := \{f\in \mathcal L^2(\Gamma): f \text{ continuously differentiable, } \tilde q^\kappa(f)<\infty\}.
\end{equation*}
We note that Assumption \ref{assump:kappa} is essential here. From a probabilistic point of view, we need $\tilde \kappa$ to be bounded below since a sensible killing rate must be nonnegative (which amounts to putting a bound on the Radon--Nikod\'{y}m derivative; see \cite[Appendix B]{Pollock2016}). From a functional-analytic point of view, we also need $\tilde \kappa$ to be bounded below since we require $\tilde q^\kappa$ to be closable. The semi-boundedness assumption on $\tilde{\kappa}$ implies that for all compactly supported, twice differentiable $f\in C^2_c(\mathbb{R}^d)$, $\tilde q^{\kappa}(f)$ is a nonnegative quadratic form associated to the symmetric operator $\tilde L^{\kappa}$. By Lemma 1.29, Assertion 2 of \cite{Ouhabaz2005} we therefore conclude that the quadratic form $\tilde q^\kappa$ is closable.

Now let us denote the closure of $\tilde q^\kappa$ by $q^\kappa$. To this quadratic form, there is associated a unique positive self-adjoint operator $L^\kappa$, with dense domain $\mathcal D(L^\kappa) \subset \mathcal L^2(\Gamma)$; see \cite[Section 1.2.3]{Ouhabaz2005}. For smooth functions the action of $L^\kappa$ is identical to that of $\tilde L^\kappa$.

Let $\Sigma(L^\kappa)$ denote the $\mathcal L^2(\Gamma)$-spectrum of $L^\kappa$. Since $L^\kappa$ is self-adjoint and positive, we have that $\Sigma(L^\kappa) \subset [0,\infty)$. We have seen in \eqref{eq:L_varphi} that $K\in\Sigma(L^\kappa)$; in particular $\Sigma(L^\kappa)$ is nonempty, so let us write $\lambda_0^\kappa$ for the bottom of the spectrum. In fact, we have that $K=\lambda_0^\kappa$. This follows from general operator theory 
since $\varphi$ is positive everywhere. We also have that $\lambda_0^\kappa$ is a simple eigenvalue, with $\varphi$ being its unique eigenfunction up to constant multiples. A reference for these assertions is \cite[Section XIII.12]{Reed1978}.

We now make use of the spectral calculus
for self-adjoint operators using projection-valued measures, as discussed in \cite[Section 2.5]{davies1995spectral}. This gives us the existence of a family of spectral projections $(E^\kappa_\lambda)_{\lambda \in [\lambda_0^\kappa,\infty)}$ and allows us to define $\phi(L^\kappa)$ for Borel-measurable $\phi: \mathbb R\rightarrow \mathbb R$, via

\begin{align*}
\phi(L^\kappa)f &= \int_{\Sigma(L^\kappa)} \phi(\lambda) \dif E^\kappa_\lambda f,
\\ \mathcal D(\phi(L^\kappa))&= \bigg \{f\in \mathcal L^2(\Gamma): \int_{\Sigma(L^\kappa)} |\phi(\lambda)|^2 \dif\langle E^\kappa_\lambda f,f\rangle_{\mathcal L^2(\Gamma)} <\infty\bigg\},
\\ \|\phi(L^\kappa) f\|_{\mathcal L^2(\Gamma)}^2 &=\int_{\Sigma(L^\kappa)} |\phi(\lambda)|^2 \dif\langle E^\kappa_\lambda f,f\rangle_{\mathcal L^2(\Gamma)}\; . 
\end{align*}
Now the Feynman--Kac representation states that for each $t>0$
\begin{equation*}
(e^{-tL^\kappa}f)(x) = \mathbb E_x[f(X_t) 1_{\{\tau_\partial > t\}}]
\end{equation*}
for $f\in \mathcal L^2(\Gamma)$. Furthermore, for each $t> 0$ the operator $e^{-tL^\kappa}$ is a contraction on $\mathcal L^2(\Gamma)$ ({\em cf.} the derivation in \cite{Demuth2000}). 

The spectral theorem allows us to write the diffusion semi-group as
\begin{equation*}
P^\kappa_t f (x) =\mathbb E_x[f(X_t) 1_{\{\tau_\partial >t\}}]=e^{-tL^\kappa}f(x)=\int_{\Sigma(L^\kappa)} e^{-t\lambda} \dif E_\lambda^\kappa f(x)
\end{equation*}
for $f\in \mathcal L^2(\Gamma)$. The $(E_\lambda^\kappa)_{\lambda \in [\lambda_0^\kappa, \infty)}$ are orthogonal projections; in particular, $E_{\lambda_0^\kappa}^\kappa$ projects onto the span of $\varphi$. 
We can write 
\begin{equation*}
e^{-tL^\kappa}f = e^{-t\lambda_0^\kappa}\varphi\langle f,\varphi \rangle_{\mathcal L^2(\Gamma)} + \int_{\Sigma(L^{\kappa}) \setminus \lbrace \lambda_0^{\kappa}\rbrace } e^{-t\lambda} \dif E_\lambda^\kappa f \; .
\end{equation*}
Thus
\begin{equation}
e^{t\lambda_0^\kappa} e^{-tL^\kappa}f = \varphi\langle f,\varphi \rangle_{\mathcal L^2(\Gamma)} + \int_{\Sigma(L^{\kappa}) \setminus \lbrace \lambda_0^{\kappa}\rbrace} e^{-t(\lambda-\lambda_0^\kappa)} \dif E_\lambda^\kappa f \; .
\label{eq:spectralrep}
\end{equation}
For a given $f\in \mathcal L^2$, we are interested in the convergence to 0  of the integral term in (\ref{eq:spectralrep}). We note here that the convergence in this discussion is convergence in $\mathcal L^2(\Gamma)$. Ultimately we will be interested in convergence in $\mathcal L^1 (\Gamma)$; we will return to this issue later.

\section{Proof of Theorem \ref{thm:qsd}}
\label{sec:proofqsd}

We wish to apply the results of \cite{Tuominen1979}. In order to do this, we first need to check that $(P^\kappa_t)_{t\ge 0}$ is ``simultaneously $\phi$-irreducible'', that is, the resolvent
kernel is strictly positive for discrete versions of the process discretised with respect to arbitrary time-steps. Ordinary $\phi$-irreducibility holds for diffusions with smooth drift and locally bounded by the Stroock--Varadhan support theorem, \cite[Section 2.6]{Pinsky1995}. Simultaneous $\phi$-irreducibility follows then immediately from Theorem 1 of \cite{Tuominen1979} since our process has a jointly continuous transition density with respect to the reversing measure; see Remark 1 after this proof.

We now show that $(P^\kappa_t)_{t\ge 0}$ is $\lambda$-positive, with $\lambda = K$, and that the $K$-invariant measure is precisely the target density $\pi$. This will then imply convergence to quasi-stationarity by an application of Theorem 7 of \cite{Tuominen1979}, which states that $\lambda$-positive processes, when $\lambda>0$, exhibit quasi-limiting convergence as in \eqref{eq:quasilimit}, where the quasi-limiting distribution is the (unique) $\lambda$-invariant measure.

By Theorem 4(ii) of \cite{Tuominen1979}, showing $(P^\kappa_t)_{t\ge 0}$ is $\lambda$-positive is equivalent to showing that each (discrete-time) skeleton chain generated by $P^\kappa_h$, for any $h>0$, is $e^{\lambda h}$-positive in the discrete-time sense, as defined in \cite{Tweedie1974}. This involves showing that each skeleton chain is $R$-recurrent with $R=e^{\lambda h}$ and that the corresponding integral of the $e^{\lambda h}$-invariant function against the $e^{\lambda h}$-invariant measure is finite. So let us fix $h>0$.

It follows from \eqref{eq:L_varphi} and the Kolmogorov equations that
\begin{equation*}
e^{hK}P^\kappa_h \varphi = \varphi.
\end{equation*}
	This is exactly the definition of $\varphi$ being $e^{hK}$-invariant for the discrete-time semi-group generated by $P^\kappa_h$. By \eqref{eq:adjointaction}
the measure $\pi$ with Lebesgue density $\gamma\varphi$
is similarly $e^{hK}$-invariant for the discrete-time chain. (Definitions of $\lambda$-invariance are included in Section \ref{sec:background_semi-group} for convenience.)

By Assumption \ref{assump:L2},
\begin{equation*}
\int_{\R^d} \varphi(y) \pi(\dif y) = \int_{\R^d} \frac{\pi^2(y)}{\gamma(y)}\dif y <\infty.
\end{equation*}
Thus by Proposition 3.1 and Proposition 4.3 of \cite{Tweedie1974} the skeleton chain defined by operator $P^\kappa_h$, $(X_{nh})_{n=1}^\infty$, is $R$-recurrent, with $R=e^{hK}$. Theorem 7 of \cite{Tweedie1974} then tells us that this skeleton chain is $e^{hK}$-positive. Since $h>0$ was arbitrary, we obtain that $(P^\kappa_t)_{t\ge 0}$ is $\lambda$-positive, with $\lambda=K$. Theorem 4(iii) of \cite{Tuominen1979} also tells us that $\varphi$ and $\pi$ are the unique $K$-invariant function and measure for $(P^\kappa_t),$ respectively.

We are now in a position to utilise Theorem 7 of \cite{Tuominen1979}. Since $K>0$, killing happens almost surely, hence the key assumption (B) of Theorem 7 of \cite{Tuominen1979} requires simply that $\int \pi(y)\dif y <\infty$, which is certainly true. The conclusion of the theorem implies convergence to quasi-stationarity \eqref{eq:quasilimit}; that is, for any measurable $E\subset \R^d$ there is a set of starting
points $x$ of full Lebesgue measure such that

$$ 
\lim_{t\rightarrow \infty} \mathbb P_x(X_t \in E \, |\tau_\partial > t) = \frac{\int_E \pi(y)\dif y}{\int_{\R^d} \pi(y) \dif y}.
$$
In fact, this convergence holds for every starting point $x$. Since we have a continuous transition density $p^\kappa(t,x,y)$ (see Remark 1 after this proof), we have for any measurable set $E\subset \Rd$
$$ 
\P_x(X_{t+1} \in E) = \int_{\Rd} p^\kappa(1,x,y)\P_y(X_t \in E)\dif\Gamma(y).
$$ 
Since we have convergence for $y$ in some set of full measure, we obtain convergence for all $x\in \Rd$, which completes the desired result. $\Box$

\vspace{4mm}

\noindent\textbf{Remarks}
\begin{enumerate}
\item Assumption \ref{assump:L2} can be interpreted in terms of spectral theory. It tells us that $\varphi \in \mathcal L^2(\mathbb R^d, \Gamma)$, so $\varphi$ is also an eigenfunction of $L^\kappa$ in the sense of $\mathcal L^2$ spectral theory. It is then possible to prove Theorem \ref{thm:qsd} analogously to Lemma 4.4 of \cite{Kolb2012}. Following the derivation of \cite{Demuth2000}, it follows that we have a continuous integral kernel $p^\kappa(t,x,y)$ with $\mathbb E_x [f(X_t) 1_{\{\tau_\partial >t\}}] = \int p^\kappa (t,x,y) f(y) \dif\Gamma(y)$. We can then apply \cite{Simon1993} to see that $e^{t\lambda^\kappa_0}p^\kappa (t,x,y) \rightarrow c \varphi(x) \varphi(y)$ as $t\rightarrow \infty$, where $c=\|\varphi\|_{\mathcal L^2(\Gamma)}^{-2}$ and the proof of Theorem \ref{thm:qsd} can proceed analogously.
\item Our argument here relies fundamentally on self-adjointness of the operators and subsequent properties such as \eqref{eq:L_varphi}
, so there is no way we can circumvent the assumption of a gradient-form drift in (\ref{eq:SDE_X}). In one dimension this always holds, since we can simply take the integral of the drift function.
\end{enumerate}

\section{Rates of convergence}
\label{sec:rates_of_conv}
Practitioners hoping to implement quasi-stationary Monte Carlo methods, such as the ScaLE Algorithm of \cite{Pollock2016}, having been reassured that the procedure indeed converges to the correct distribution, will naturally inquire about the rate of convergence. Our result in this section draws heavily on the spectral theory for self-adjoint (unbounded) operators that we have outlined in Section \ref{sec:background_operator}.

When there is a \textit{spectral gap}, that is, when $\lambda_1^\kappa > \lambda_0^\kappa$, the integral term will vanish at an exponential rate.
Thus, it suffices to understand the spectrum $\Sigma(L^\kappa)$. To do this, we will adapt an idea of \cite{Pinsky2009}, to transform our operator into one whose spectrum is already understood. Here it will
be the infinitesimal generator of a certain Langevin diffusion.

\subsection{Proof of Theorem \ref{thm:rate_conv}}
\label{subsec:pf_thm2}
Consider the formal differential operator
\begin{equation*}
\tilde L^{\tilde \kappa} = -\frac{1}{2\gamma}\nabla \cdot\gamma  \nabla + \tilde \kappa
\end{equation*}
where $\tilde \kappa$ is defined in (\ref{eq:kappa_tilde}), acting on $C_c^\infty (\mathbb R^d)$, the set of smooth compactly-supported functions. This is very similar to the formal differential operator we began with in (\ref{eq:tilde_L_kappa}), differing only by an additive constant $K$, which will have the effect of merely translating the spectrum accordingly. $\tilde L^{\tilde \kappa}$ can be realised as a nonnegative, self-adjoint operator $L^{\tilde \kappa}$ on $\mathcal L^2(\Gamma)$, by taking the Friedrichs extension of the appropriate quadratic form as before.

Now let $\mathcal L^2(\pi^2/\gamma)\equiv\mathcal L^2(\mathbb R^d, \pi^2/\gamma)$ denote the Hilbert space of (equivalence classes of) measurable functions $u,v: \mathbb R^d\rightarrow \mathbb R$ which are square-integrable with respect to the inner product
\begin{equation*}
\langle u,v\rangle_{\mathcal L^2(\pi^2/\gamma)} = \int_{\mathbb R^d} u(y)v(y) \frac{\pi^2(y)}{\gamma(y)}\dif y.
\end{equation*}
The multiplication operator
\begin{equation*}
Uf = \frac{\gamma}{\pi} f
\end{equation*}
is a bounded unitary transformation $U:\mathcal L^2(\Gamma)\rightarrow \mathcal L^2(\pi^2/\gamma)$, with 
inverse given by $U^{-1} u = \frac{\pi}{\gamma}u$.

We now define a second formal differential operator
\begin{equation*}
\tilde L^{Z} = -\frac{1}{2}\Delta -\frac{1}{2}\nabla \log\bigg(\frac{\pi^2}{\gamma}\bigg) \cdot \nabla\; ,
\end{equation*}
which is minus the generator of the Langevin diffusion given in (\ref{eq:Lang_pi2gamma}), targeting the density $\pi^2/\gamma$. $\tilde L^{Z}$ can similarly be realized as a positive, self-adjoint operator $L^{Z}$ on $\mathcal L^2(\pi^2/\gamma)$.
Our two formal operators are related through
\begin{equation*}
\tilde L^{\tilde \kappa} = U^{-1} \tilde L^{Z} U.
\end{equation*}
We can also conjugate $L^{Z}$ to obtain $U^{-1} L^{Z} U$, a self-adjoint operator on $\mathcal L^2(\Gamma)$. 

Theorem \ref{thm:rate_conv} will be an immediate consequence if we show that in fact $U^{-1} L^{Z} U=L^{\tilde \kappa}$. This is the same as showing that the following diagram commutes:

\begin{equation*}
\begin{tikzcd}[column sep = huge, row sep = huge]
\tilde L^{Z} \arrow{r}{\text{Friedrichs ext.}} \arrow[swap]{d}{\text{Conjugate with }U} & L^{Z}  \arrow{d}{\text{Conjugate with }U} \\%
\tilde L^{\tilde \kappa} \arrow{r}{\text{Friedrichs ext.}}& L^{\tilde \kappa}
\end{tikzcd}
\end{equation*}
An operator is said to be {\em essentially self-adjoint} if it has a unique self-adjoint extension, which is given by the closure. 
From the background in Section \ref{sec:background_operator}, we see that the diagram commutes, and so Theorem \ref{thm:rate_conv}
will follow, if we can show that $\tilde L^{\tilde \kappa}$ acting on $C_c^\infty (\mathbb R^d)$ is essentially self-adjoint.
After all, the conjugate $U^{-1} \tilde L^{Z} U$
is a self-adjoint extension of $\tilde L^{\tilde \kappa}$; if the extension is unique it must be the same as $L^{\tilde\kappa}$.

We apply Theorem 2.13 of \cite{Braverman2002a}. The smooth boundaryless manifold we are working in is simply $\mathbb R^d$, with smooth positive measure measure $\Gamma$. In their notation, we take $D$ to be $\frac{1}{\sqrt 2}\nabla$, which is elliptic. The formal adjoint $D^*$ is given by $-\frac{1}{\sqrt 2}(\nabla \cdot +2\nabla A\cdot)$. We set $V=\tilde \kappa$, and the resulting operator $H_V$ is precisely $\tilde L^{\tilde \kappa}$.

The result follows immediately if $V$ satisfies their Assumptions A and B, which ask for a decomposition of $V$ into well-behaved nonnegative parts and a mild technical condition. Assumption A is immediate by writing
\begin{equation*}
V= \underbrace{\tilde \kappa +K}_{V_+}  +\underbrace{(-K)}_{V_-}
\end{equation*}
where clearly $V_+ \ge 0$ and $V_- \le 0$. $V_-$ trivially satisfies (ii) of Assumption A since it is constant.

Assumption B follows from their Theorem 2.3(ii), since our operator acts on scalar functions. The final condition of Theorem 2.13 is completeness of the metric $g^{TM}$, which is satisfied since it is equivalent to geodesic completeness of the manifold, which is true for $\mathbb R^d$.

Since unitary transformations leave spectra invariant it follows that the $\mathcal L^2(\Gamma)$ spectrum of $L^{\tilde \kappa}$ coincides with the $\mathcal L^2(\pi^2/\gamma)$ spectrum of $L^Z$, and hence the $\mathcal L^2$ spectra of $L^{\tilde \kappa}$ and $L^Z$ coincide after translation by $K$.

\renewcommand{\L}{\mathcal{L}}
We now would like to extend our proof of $\L^2$ convergence to $\L^1$ convergence in the case when there is a spectral gap. 
Let $\psi \in \L^1(\Gamma)\cap \L^2(\Gamma)$ be any initial density (with respect to the measure $\Gamma$). For the rest of this section, all norms and inner products will be with respect to $\L^2(\Gamma)$.
Writing $\lambda_1^\kappa:= \inf \big \{\Sigma(L^\kappa)\setminus \{\lambda_0^\kappa\}\big \}$, from our earlier results we have that
\begin{align}
\|e^{t\lambda_0^\kappa} e^{-tL^\kappa} \psi - \langle\psi, \varphi \rangle\varphi \|^2 &=\bigg \|\int_{\lambda_1^\kappa}^\infty e^{-t(\lambda-\lambda_0^\kappa)} \dif E_\lambda^\kappa \psi\bigg \|^2 \notag \\
&=\int_{\lambda_1^\kappa}^\infty e^{-t\cdot 2(\lambda-\lambda_0^\kappa) } \dif\langle E_\lambda^\kappa \psi,\psi\rangle \notag\\
&\le \|\psi\|^2 \cdot e^{-t\cdot 2(\lambda_1^\kappa - \lambda_0^\kappa)}.
\label{eq:convergeL2}
\end{align}

We now link this to $\L ^1$ convergence. Let $H\subset \mathbb R^d$ be a compact set. 
From the Cauchy--Schwarz inequality, we know that
\begin{align*}
\int_H |e^{t\lambda_0^\kappa} e^{-tL^\kappa}&\psi(y) - \langle \psi, \varphi \rangle \varphi (y)|\,\dif\Gamma(y)\\ 
&\le \|e^{t\lambda_0^\kappa} e^{-tL^\kappa} \psi - \langle\psi, \varphi \rangle\varphi \|
\cdot \Gamma(H)^{1/2}\\
&\le \|\psi\| \cdot \Gamma(H)^{1/2}\cdot  e^{-t(\lambda_1^\kappa-\lambda_0^\kappa)}.
\end{align*}
So we have the appropriate convergence in $\L^1(\Gamma)$ on compact sets. We could similarly obtain convergence for test functions $f\in \L^2(\Gamma)$, that is,
\begin{equation} \label{eq:convergeL1}
\Bigl|\bigl\langle e^{t\lambda_0^\kappa} e^{-tL^\kappa} \psi,f \bigr\rangle -  \langle \psi, \varphi\rangle \langle\varphi,f \rangle
\Bigr| \le \|\psi\|\cdot \|f\| \cdot e^{-t(\lambda_1^\kappa-\lambda_0^\kappa)}.
\end{equation}
We see that when $\Gamma$ is a finite measure, we will obtain $\L^1$ convergence at this rate on all measurable sets, not just compact ones. This is the case when the (unkilled) diffusion has a strong inward drift.

Now assume that $\Gamma(\R^d)<\infty$ and fix some $E\subset \mathbb R^d$. Writing $\mathbb P_\psi$ for the law of the killed process starting from $\psi$, we have (recalling that $\int_{\R^d} \pi(x)\dif x =\langle \varphi,1\rangle$),
\begin{align*}
\bigl|\mathbb P_\psi &(X\in E\,  | \,\tau_\partial >t) -\pi(E)\bigr|=\bigg |\int_E \bigg(\frac{e^{t\lambda_0^\kappa} e^{-tL^\kappa}\psi(y)}{e^{t\lambda_0^\kappa}\mathbb P_\psi (\tau_\partial >t)} -\frac{\varphi(y)}{\int_{\mathbb R^d} \pi(x)\dif x}\bigg)\,\dif\Gamma(y)\bigg |\\ 
&= \frac{1}{\langle \psi, \varphi \rangle\int \pi(x)dx}\bigg |\int_E \bigg(e^{t\lambda_0^\kappa} e^{-tL^\kappa}\psi(y) 
   -\langle \psi, \varphi \rangle\varphi(y)\bigg)\,\dif\Gamma(y)\\
&\hspace*{0.75cm} -\biggl(\int_E e^{t\lambda_0^\kappa} e^{-tL^\kappa}\psi(y) \dif\Gamma(y) \biggr)
\biggl(\frac{e^{t\lambda_0^\kappa}\mathbb P_\psi (\tau_\partial >t)-\langle \psi, \varphi \rangle\int \pi(x)dx}{e^{t\lambda_0^\kappa}\mathbb P_\psi (\tau_\partial >t)}\biggr)
\bigg|.
\end{align*}
Note that
$$
\frac{\int_E e^{t\lambda_0^\kappa} e^{-tL^\kappa}\psi(y) \dif\Gamma(y)}{e^{t\lambda_0^\kappa}\mathbb P_\psi (\tau_\partial >t)}= \frac{\langle e^{t\lambda_0^\kappa} e^{-tL^\kappa}\psi,\mathbf{1}_E\rangle}{\langle e^{t\lambda_0^\kappa} e^{-tL^\kappa}\psi,\mathbf{1}\rangle}\le 1,
$$
so
\begin{equation} \label{eq:convQSDrate}
\bigl|\mathbb P_\psi (X\in E \, |\, \tau_\partial >t) -\pi(E)\bigr|
\le 
\frac{2\|\psi\|\Gamma\left( \R^d \right)^{1/2}}{\langle \psi, \varphi \rangle\int \pi(x)dx} e^{-t(\lambda_1^\kappa - \lambda_0^\kappa)}.
\end{equation}

It remains to derive the rate of pointwise convergence for 
$$
e^{t\lambda_0^\kappa} p^\kappa (t,x,y)\rightarrow  \varphi (x) \varphi (y) \quad \text{ as }t\rightarrow \infty.
$$
This argument does not require us to assume $\Gamma(\Rd)<\infty$. Following the approach of \cite{Simon1993}, for $x,y\in \R^d$ let us write $g_x(y):= e^{\lambda_0^\kappa} p^\kappa(1,x,y)$. First, note that $g_x \in \mathcal L^2(\Gamma)$:
\begin{align*}
\|g_x\|_{\mathcal L^2(\Gamma)}^2 &=e^{2\lambda_0^\kappa}\int p^\kappa (1,x,y) p^\kappa (1,x,y) \,\dif\Gamma(y)\\
&=e^{2\lambda_0^\kappa} \int p^\kappa(1,x,y)p^\kappa(1,y,x)\,\dif\Gamma(y)\\
&=e^{2\lambda_0^\kappa} p^\kappa(2,x,x)\\
&< \infty,
\end{align*}
using symmetry and the semi-group property. By the invariance of $\varphi$,
$$
\langle g_x,\varphi\rangle = e^{\lambda_0^\kappa}\int p^\kappa (1,x,z)\varphi(z) \dif\Gamma(z)=\varphi(x).
$$

Now for $t>2, x, y \in \R^d,$
\begin{align*}
e^{t\lambda_0^\kappa}p^\kappa (t, x,y)&=\int \int g_x(z) e^{\lambda_0^\kappa (t-2)}p^\kappa(t-2,z,w) g_y (w) \,\dif\Gamma( z)\dif\Gamma( w)\\
&= \bigl\langle e^{-(t-2)L^\kappa}e^{(t-2)\lambda_0^\kappa }g_x \, , \, g_y \bigr\rangle.
\end{align*}
By \eqref{eq:convergeL1}, this converges to
$$
\langle g_x,\varphi\rangle\langle g_y,\varphi\rangle
 = \varphi(x)\varphi(y),
$$
with rate given by
\begin{equation} \label{eq:convkernel}
\bigl| e^{t\lambda_0^\kappa}p^\kappa (t, x,y) - \varphi(x)\varphi(y) \bigr|\le
  e^{2\lambda_0^\kappa} \bigl(p^{\kappa}(2,x,x)
    p^{\kappa}(2,y,y) \bigr)^{1/2} e^{-t(\lambda_1^\kappa-\lambda_0^\kappa)}.
\end{equation}
If the drift is bounded then the transition density is bounded as well, so this is bounded by $Ce^{-t(\lambda_1^\kappa-\lambda_0^\kappa)}$ for a universal constant $C$. $\Box$

\section{Discussion}

In this paper, we have proven natural sufficient conditions for the quasi-limiting distribution of a diffusion of the form \eqref{eq:SDE_X} killed at an appropriate state-dependent rate to coincide with a target density $\pi$. We have also quantified the rate of convergence to quasi-stationarity by relating the rate of this convergence to the rate of convergence to stationarity of a related unkilled process.

As mentioned in the Introduction, this framework is foundational for the recently-developed class of quasi-stationary Monte Carlo algorithms to sample from Bayesian posterior distributions, introduced in \cite{Pollock2016}. This framework promises improvement over  more traditional MCMC approaches particularly for Bayesian inference on large datasets, since the killed diffusion framework enables the use of \textit{subsampling} techniques. As detailed in \cite[Section 4]{Pollock2016}, these allow the construction of estimators which scale exceptionally well as the size of the underlying dataset grows.

Quasi-stationary Monte Carlo methods are likely to be particularly effective compared to established Monte Carlo methods for Bayesian inference for {\em tall data}; that is, where parameter spaces have moderate dimension (allowing diffusion simulation to be feasible) but where data sizes are high. This includes the `Big-Data' context where data size is so large it cannot even be stored locally on computers implementing the algorithm. This is because subsampling can take place `offline' with only the subsets being stored locally.  This adds significantly to the potential applicability of quasi-stationary Monte Carlo methods.

Our approach in this present work is also slightly more general than that of \cite{Pollock2016} in that we allow for a nonzero drift term in our diffusion \eqref{eq:SDE_X}. This raises the question of how to select among several possible drift functions the one that results in the most practical computational outcomes. While a detailed answer to this question is beyond the scope of this present work, we suggest the following guidelines. There is a critical trade-off between overall killing and the essential
rate of convergence described in \eqref{eq:quasilimit}. As mentioned in the example of Section \ref{sec:OU_example}, higher rates of killing will tend to \textit{increase} the essential rate of convergence, while increasing the computational burden imposed by simulating killing events. Depending on the details of the implementation, this trade-off could go either way in terms of optimality. When scalable estimators for the killing events are available, such as in \cite{Pollock2016}, it would be sensible to choose a drift that makes the killing rate high, for instance choosing a Brownian motion, so $A\equiv 0$. Of course, any Gaussian process allows straightforward simulation of the unkilled dynamics, and the choice of Brownian motion also simplifies Assumptions \ref{assump:L2} and \ref{assump:kappa}. Formally answering this question of the choice of drift would be an interesting avenue for future exploration.

We comment briefly now on some of our assumptions. Assumption \ref{assump:L2} is generally straightforward to verify, especially in light of the stronger `rejection sampling' formulation in \eqref{eq:reject_sampl}. For instance, if $A$ is uniformly bounded below then Assumption \ref{assump:L2} holds if $\pi$ is a bounded density function.

Assumption \ref{assump:kappa} is generally the most challenging. When $A\equiv 0,$ this is mostly straightforward to verify, especially since densities on $\Rd$ are often convex in the tails. Verifying Assumption \ref{assump:kappa} in general can be done using the equivalent expression for $\tilde \kappa$ in \eqref{eq:tkappa_logs}, by comparing the decay of derivatives $2A$ with those of $\log\pi$. Indeed, ensuring a practically useful form of $\tilde \kappa$---so that verification of Assumption \ref{assump:kappa} is straightforward---could influence the choice of $A$ in the first place.

In practice, Assumption \ref{assump:kappa} also involves computing a lower bound for $\tilde \kappa$. It is actually not necessary to compute the precise value of $\inf_{y\in\Rd} \tilde \kappa(y)$; our results still hold if $K$ in \eqref{eq:kappa} is replaced by any constant such that the resulting $\kappa$ is nonnegative everywhere. 
Intuitively, taking a larger constant $K$ amounts to merely adding additional killing events according to a homogeneous, independent Poisson process. 

Depending on the choices of $\pi$ and $A$, $\tilde \kappa$ can be a convex function in the tails, even in cases of nonzero $A$, as in our example of Section \ref{sec:OU_example}. 
A precise recipe for computing $K$ in general is currently unavailable; readers interested in these more implementational details are encouraged to look at \cite{Pollock2016}.

We conclude this discussion by indicating some potential future directions. As mentioned above, there are important questions of how to choose the underlying diffusion to optimize the computation for a given target density. One could also consider extensions of this work to entirely different underlying processes, such as jump diffusions or L\'{e}vy processes. Finally, another potential question is the exploration of alternative approaches to that described in \cite{Pollock2016} for the simulation of quasi-stationary distributions, such as the stochastic approximation approaches as discussed in \cite{Blanchet2016} and \cite{Benaim16}.

\bibliography{QSMC_bib}

\begin{thebibliography}{26}

\bibitem[\protect\citeauthoryear{Bardenet, Doucet and
  Holmes}{2017}]{Bardenet17}
\begin{barticle}[author]
\bauthor{\bsnm{Bardenet},~\bfnm{R\'emi}\binits{R.}},
  \bauthor{\bsnm{Doucet},~\bfnm{Arnaud}\binits{A.}} \AND
  \bauthor{\bsnm{Holmes},~\bfnm{Chris}\binits{C.}}
(\byear{2017}).
\btitle{On {M}arkov chain {M}onte {C}arlo methods for tall data}.
\bjournal{J. Mach. Learn. Res.}
\bvolume{18}
\bpages{Paper No. 47, 43}.
\bmrnumber{3670492}
\end{barticle}
\endbibitem

\bibitem[\protect\citeauthoryear{Bena{\"{i}}m, Cloez and
  Panloup}{2018}]{Benaim16}
\begin{barticle}[author]
\bauthor{\bsnm{Bena{\"{i}}m},~\bfnm{Michel}\binits{M.}},
  \bauthor{\bsnm{Cloez},~\bfnm{Bertrand}\binits{B.}} \AND
  \bauthor{\bsnm{Panloup},~\bfnm{Fabien}\binits{F.}}
(\byear{2018}).
\btitle{Stochastic approximation of quasi-stationary distributions on compact
  spaces and applications}.
\bjournal{Ann. Appl. Probab.}
\bvolume{28}
\bpages{2370--2416}.
\bdoi{10.1214/17-AAP1360}
\bmrnumber{3843832}
\end{barticle}
\endbibitem

\bibitem[\protect\citeauthoryear{Beskos, Papaspiliopoulos and
  Roberts}{2006}]{Beskos2006}
\begin{barticle}[author]
\bauthor{\bsnm{Beskos},~\bfnm{Alexandros}\binits{A.}},
  \bauthor{\bsnm{Papaspiliopoulos},~\bfnm{Omiros}\binits{O.}} \AND
  \bauthor{\bsnm{Roberts},~\bfnm{Gareth~O.}\binits{G.~O.}}
(\byear{2006}).
\btitle{Retrospective exact simulation of diffusion sample paths with
  applications}.
\bjournal{Bernoulli}
\bvolume{12}
\bpages{1077--1098}.
\bdoi{10.3150/bj/1165269151}
\bmrnumber{2274855}
\end{barticle}
\endbibitem

\bibitem[\protect\citeauthoryear{Blanchet, Glynn and
  Zheng}{2016}]{Blanchet2016}
\begin{barticle}[author]
\bauthor{\bsnm{Blanchet},~\bfnm{J.}\binits{J.}},
  \bauthor{\bsnm{Glynn},~\bfnm{P.}\binits{P.}} \AND
  \bauthor{\bsnm{Zheng},~\bfnm{S.}\binits{S.}}
(\byear{2016}).
\btitle{Analysis of a stochastic approximation algorithm for computing
  quasi-stationary distributions}.
\bjournal{Adv. in Appl. Probab.}
\bvolume{48}
\bpages{792--811}.
\bdoi{10.1017/apr.2016.28}
\bmrnumber{3568892}
\end{barticle}
\endbibitem

\bibitem[\protect\citeauthoryear{Braverman, Milatovich and
  Shubin}{2002}]{Braverman2002a}
\begin{barticle}[author]
\bauthor{\bsnm{Braverman},~\bfnm{M.}\binits{M.}},
  \bauthor{\bsnm{Milatovich},~\bfnm{O.}\binits{O.}} \AND
  \bauthor{\bsnm{Shubin},~\bfnm{M.}\binits{M.}}
(\byear{2002}).
\btitle{Essential selfadjointness of {S}chr\"odinger-type operators on
  manifolds}.
\bjournal{Uspekhi Mat. Nauk}
\bvolume{57}
\bpages{3--58}.
\bdoi{10.1070/RM2002v057n04ABEH000532}
\bmrnumber{1942115}
\end{barticle}
\endbibitem

\bibitem[\protect\citeauthoryear{Champagnat and
  Villemonais}{2016}]{Champagnat2016a}
\begin{barticle}[author]
\bauthor{\bsnm{Champagnat},~\bfnm{Nicolas}\binits{N.}} \AND
  \bauthor{\bsnm{Villemonais},~\bfnm{Denis}\binits{D.}}
(\byear{2016}).
\btitle{Exponential convergence to quasi-stationary distribution and
  {$Q$}-process}.
\bjournal{Probab. Theory Related Fields}
\bvolume{164}
\bpages{243--283}.
\bdoi{10.1007/s00440-014-0611-7}
\bmrnumber{3449390}
\end{barticle}
\endbibitem

\bibitem[\protect\citeauthoryear{Collet, Mart\'{i}nez and
  San~Mart\'{i}n}{2013}]{Collet2012}
\begin{bbook}[author]
\bauthor{\bsnm{Collet},~\bfnm{Pierre}\binits{P.}},
  \bauthor{\bsnm{Mart\'{i}nez},~\bfnm{Servet}\binits{S.}} \AND
  \bauthor{\bsnm{San~Mart\'{i}n},~\bfnm{Jaime}\binits{J.}}
(\byear{2013}).
\btitle{{Quasi-stationary distributions: Markov chains, Diffusions and
  Dynamical Systems}}.
\bpublisher{Springer, Heidelberg}.
\bdoi{10.1007/978-3-642-33131-2}
\bmrnumber{2986807}
\end{bbook}
\endbibitem

\bibitem[\protect\citeauthoryear{Dalalyan}{2017}]{Dalalyan2016}
\begin{barticle}[author]
\bauthor{\bsnm{Dalalyan},~\bfnm{Arnak~S.}\binits{A.~S.}}
(\byear{2017}).
\btitle{Theoretical guarantees for approximate sampling from smooth and
  log-concave densities}.
\bjournal{J. R. Stat. Soc. Ser. B. Stat. Methodol.}
\bvolume{79}
\bpages{651--676}.
\bdoi{10.1111/rssb.12183}
\bmrnumber{3641401}
\end{barticle}
\endbibitem

\bibitem[\protect\citeauthoryear{Davies}{1995}]{davies1995spectral}
\begin{bbook}[author]
\bauthor{\bsnm{Davies},~\bfnm{E.~B.}\binits{E.~B.}}
(\byear{1995}).
\btitle{{Spectral Theory and Differential Operators}}.
\bseries{Cambridge Studies in Advanced Mathematics}
\bvolume{42}.
\bpublisher{Cambridge Univ. Press, Cambridge}.
\bdoi{10.1017/CBO9780511623721}
\bmrnumber{1349825}
\end{bbook}
\endbibitem

\bibitem[\protect\citeauthoryear{Demuth and van Casteren}{2000}]{Demuth2000}
\begin{bbook}[author]
\bauthor{\bsnm{Demuth},~\bfnm{Michael}\binits{M.}} \AND
  \bauthor{\bparticle{van} \bsnm{Casteren},~\bfnm{Jan~A.}\binits{J.~A.}}
(\byear{2000}).
\btitle{{Stochastic Spectral Theory for Selfadjoint {F}eller Operators: A
  functional Integration Approach}}.
\bseries{Probability and its Applications}.
\bpublisher{Birkh\"auser Verlag, Basel}.
\bdoi{10.1007/978-3-0348-8460-0}
\bmrnumber{1772266}
\end{bbook}
\endbibitem

\bibitem[\protect\citeauthoryear{Diaconis and Miclo}{2015}]{Diaconis2014}
\begin{barticle}[author]
\bauthor{\bsnm{Diaconis},~\bfnm{Persi}\binits{P.}} \AND
  \bauthor{\bsnm{Miclo},~\bfnm{Laurent}\binits{L.}}
(\byear{2015}).
\btitle{On quantitative convergence to quasi-stationarity}.
\bjournal{Ann. Fac. Sci. Toulouse Math. (6)}
\bvolume{24}
\bpages{973--1016}.
\bdoi{10.5802/afst.1472}
\bmrnumber{3434264}
\end{barticle}
\endbibitem

\bibitem[\protect\citeauthoryear{Durmus and Moulines}{2017}]{Durmus2015}
\begin{barticle}[author]
\bauthor{\bsnm{Durmus},~\bfnm{Alain}\binits{A.}} \AND
  \bauthor{\bsnm{Moulines},~\bfnm{\'Eric}\binits{E.}}
(\byear{2017}).
\btitle{Nonasymptotic convergence analysis for the unadjusted {L}angevin
  algorithm}.
\bjournal{Ann. Appl. Probab.}
\bvolume{27}
\bpages{1551--1587}.
\bdoi{10.1214/16-AAP1238}
\bmrnumber{3678479}
\end{barticle}
\endbibitem

\bibitem[\protect\citeauthoryear{Kolb and Steinsaltz}{2012}]{Kolb2012}
\begin{barticle}[author]
\bauthor{\bsnm{Kolb},~\bfnm{Martin}\binits{M.}} \AND
  \bauthor{\bsnm{Steinsaltz},~\bfnm{David}\binits{D.}}
(\byear{2012}).
\btitle{Quasilimiting behavior for one-dimensional diffusions with killing}.
\bjournal{Ann. Probab.}
\bvolume{40}
\bpages{162--212}.
\bdoi{10.1214/10-AOP623}
\bmrnumber{2917771}
\end{barticle}
\endbibitem

\bibitem[\protect\citeauthoryear{Mandl}{1961}]{Mandl1961}
\begin{barticle}[author]
\bauthor{\bsnm{Mandl},~\bfnm{Petr}\binits{P.}}
(\byear{1961}).
\btitle{Spectral theory of semi-groups connected with diffusion processes and
  its application}.
\bjournal{Czechoslovak Math. J.}
\bvolume{11 (86)}
\bpages{558--569}.
\bmrnumber{0137143}
\end{barticle}
\endbibitem

\bibitem[\protect\citeauthoryear{Metafune, Pallara and
  Priola}{2002}]{Metafune2002}
\begin{barticle}[author]
\bauthor{\bsnm{Metafune},~\bfnm{G.}\binits{G.}},
  \bauthor{\bsnm{Pallara},~\bfnm{D.}\binits{D.}} \AND
  \bauthor{\bsnm{Priola},~\bfnm{E.}\binits{E.}}
(\byear{2002}).
\btitle{Spectrum of {O}rnstein-{U}hlenbeck operators in {$L^p$} spaces with
  respect to invariant measures}.
\bjournal{J. Funct. Anal.}
\bvolume{196}
\bpages{40--60}.
\bdoi{10.1006/jfan.2002.3978}
\bmrnumber{1941990}
\end{barticle}
\endbibitem

\bibitem[\protect\citeauthoryear{Ouhabaz}{2005}]{Ouhabaz2005}
\begin{bbook}[author]
\bauthor{\bsnm{Ouhabaz},~\bfnm{El~Maati}\binits{E.~M.}}
(\byear{2005}).
\btitle{{Analysis of Heat Equations on Domains}}.
\bseries{London Mathematical Society Monographs Series}
\bvolume{31}.
\bpublisher{Princeton Univ. Press, Princeton, NJ}.
\bmrnumber{2124040}
\end{bbook}
\endbibitem

\bibitem[\protect\citeauthoryear{Pinsky}{1995}]{Pinsky1995}
\begin{bbook}[author]
\bauthor{\bsnm{Pinsky},~\bfnm{Ross~G.}\binits{R.~G.}}
(\byear{1995}).
\btitle{{Positive Harmonic Functions and Diffusion}}.
\bseries{Cambridge Studies in Advanced Mathematics}
\bvolume{45}.
\bpublisher{Cambridge Univ. Press, Cambridge}.
\bdoi{10.1017/CBO9780511526244}
\bmrnumber{1326606}
\end{bbook}
\endbibitem

\bibitem[\protect\citeauthoryear{Pinsky}{2009}]{Pinsky2009}
\begin{barticle}[author]
\bauthor{\bsnm{Pinsky},~\bfnm{Ross~G.}\binits{R.~G.}}
(\byear{2009}).
\btitle{Explicit and almost explicit spectral calculations for diffusion
  operators}.
\bjournal{J. Funct. Anal.}
\bvolume{256}
\bpages{3279--3312}.
\bdoi{10.1016/j.jfa.2008.08.012}
\bmrnumber{2504526}
\end{barticle}
\endbibitem

\bibitem[\protect\citeauthoryear{Pollett}{2015}]{Pollett}
\begin{bmisc}[author]
\bauthor{\bsnm{Pollett},~\bfnm{Phil~K.}\binits{P.~K.}}
(\byear{2015}).
\btitle{{Quasi-stationary distributions: A bibliography}}.
\bhowpublished{http://www.maths.uq.edu.au/{\~{}}pkp/papers/qsds/qsds.html}.
\end{bmisc}
\endbibitem

\bibitem[\protect\citeauthoryear{Pollock et~al.}{2016}]{Pollock2016}
\begin{bmisc}[author]
\bauthor{\bsnm{Pollock},~\bfnm{Murray}\binits{M.}},
  \bauthor{\bsnm{Fearnhead},~\bfnm{Paul}\binits{P.}},
  \bauthor{\bsnm{Johansen},~\bfnm{Adam~M.}\binits{A.~M.}} \AND
  \bauthor{\bsnm{Roberts},~\bfnm{Gareth~O.}\binits{G.~O.}}
(\byear{2016}).
\btitle{{The Scalable Langevin Exact Algorithm: Bayesian inference for big
  data}}.
\bnote{Preprint. Available at arXiv:1609.03436.}
\end{bmisc}
\endbibitem

\bibitem[\protect\citeauthoryear{Reed and Simon}{1978}]{Reed1978}
\begin{bbook}[author]
\bauthor{\bsnm{Reed},~\bfnm{Michael}\binits{M.}} \AND
  \bauthor{\bsnm{Simon},~\bfnm{Barry}\binits{B.}}
(\byear{1978}).
\btitle{{Methods of Modern Mathematical Physics. {IV}. {A}nalysis of
  Operators}}.
\bpublisher{Academic Press, New York}.
\bmrnumber{0493421}
\end{bbook}
\endbibitem

\bibitem[\protect\citeauthoryear{Roberts and Tweedie}{1996}]{Roberts1996}
\begin{barticle}[author]
\bauthor{\bsnm{Roberts},~\bfnm{Gareth~O.}\binits{G.~O.}} \AND
  \bauthor{\bsnm{Tweedie},~\bfnm{Richard~L.}\binits{R.~L.}}
(\byear{1996}).
\btitle{Exponential convergence of {L}angevin distributions and their discrete
  approximations}.
\bjournal{Bernoulli}
\bvolume{2}
\bpages{341--363}.
\bdoi{10.2307/3318418}
\bmrnumber{1440273}
\end{barticle}
\endbibitem

\bibitem[\protect\citeauthoryear{Simon}{1993}]{Simon1993}
\begin{barticle}[author]
\bauthor{\bsnm{Simon},~\bfnm{Barry}\binits{B.}}
(\byear{1993}).
\btitle{Large time behavior of the heat kernel: on a theorem of {C}havel and
  {K}arp}.
\bjournal{Proc. Amer. Math. Soc.}
\bvolume{118}
\bpages{513--514}.
\bdoi{10.2307/2160331}
\bmrnumber{1139473}
\end{barticle}
\endbibitem

\bibitem[\protect\citeauthoryear{Tuominen and Tweedie}{1979}]{Tuominen1979}
\begin{barticle}[author]
\bauthor{\bsnm{Tuominen},~\bfnm{Pekka}\binits{P.}} \AND
  \bauthor{\bsnm{Tweedie},~\bfnm{Richard~L.}\binits{R.~L.}}
(\byear{1979}).
\btitle{Exponential decay and ergodicity of general {M}arkov processes and
  their discrete skeletons}.
\bjournal{Adv. in Appl. Probab.}
\bvolume{11}
\bpages{784--803}.
\bdoi{10.2307/1426859}
\bmrnumber{544195}
\end{barticle}
\endbibitem

\bibitem[\protect\citeauthoryear{Tweedie}{1974}]{Tweedie1974}
\begin{barticle}[author]
\bauthor{\bsnm{Tweedie},~\bfnm{Richard~L.}\binits{R.~L.}}
(\byear{1974}).
\btitle{{$R$}-theory for {M}arkov chains on a general state space. {I}.
  {S}olidarity properties and {$R$}-recurrent chains}.
\bjournal{Ann. Probability}
\bvolume{2}
\bpages{840--864}.
\bmrnumber{0368151}
\end{barticle}
\endbibitem

\bibitem[\protect\citeauthoryear{van~der Vaart}{1998}]{Vaart2000}
\begin{bbook}[author]
\bauthor{\bparticle{van~der} \bsnm{Vaart},~\bfnm{A.~W.}\binits{A.~W.}}
(\byear{1998}).
\btitle{Asymptotic Statistics}.
\bseries{Cambridge Series in Statistical and Probabilistic Mathematics}
\bvolume{3}.
\bpublisher{Cambridge Univ. Press, Cambridge}.
\bdoi{10.1017/CBO9780511802256}
\bmrnumber{1652247}
\end{bbook}
\endbibitem

\end{thebibliography}
\bibliographystyle{imsart-nameyear}
\end{document}